# The Jefferson Lab 12 GeV Program

**Hugh E. Montgomery**[1]

*Jefferson Lab*
*12000 Jefferson Avenue, Newport News, VA 23606, USA*
*E-mail:* `mont@jlab.org`

The Jefferson Lab facilities have undergone a substantial upgrade, both of accelerator, CEBAF, and of the experimental installations. We will discuss the progress to completion of these facilities, the status of accelerator commissioning, and the recent first operations for physics. We will then flesh out the anticipated exciting physics program of the next several years.

---

[1]Speaker

**1.    12 GeV CEBAF**

The CEBAF 12 GeV Upgrade project provides much of the basis for the 12 GeV program [1]. The project scope included doubling the energy of the accelerator, creating a new experimental hall, Hall D, and substantially upgrading the experimental facilities in Hall B and Hall C. In Hall A, beyond the formal project scope, a new Super BigBite Spectrometer is under construction and there are plans for future MIE projects which will supplement the two large spectrometers which will continue to function.

The accelerator upgrade within the original footprint was enabled by the enormous advances in super-conducting radio frequency acceleration since the construction CEBAF in the early 1990's. Ten new cryomodules provided the same acceleration as the original forty cryomodules. In order to deliver 12 GeV electrons to the Hall D radiator and tagger, where the polarized photon beam needed for the experiments is generated, a tenth magnetc arc was constructed and all the nine pre-existing arc magnets were refurbished. The cryogenic plant supporting the superconducting cavities was doubled in capacity.

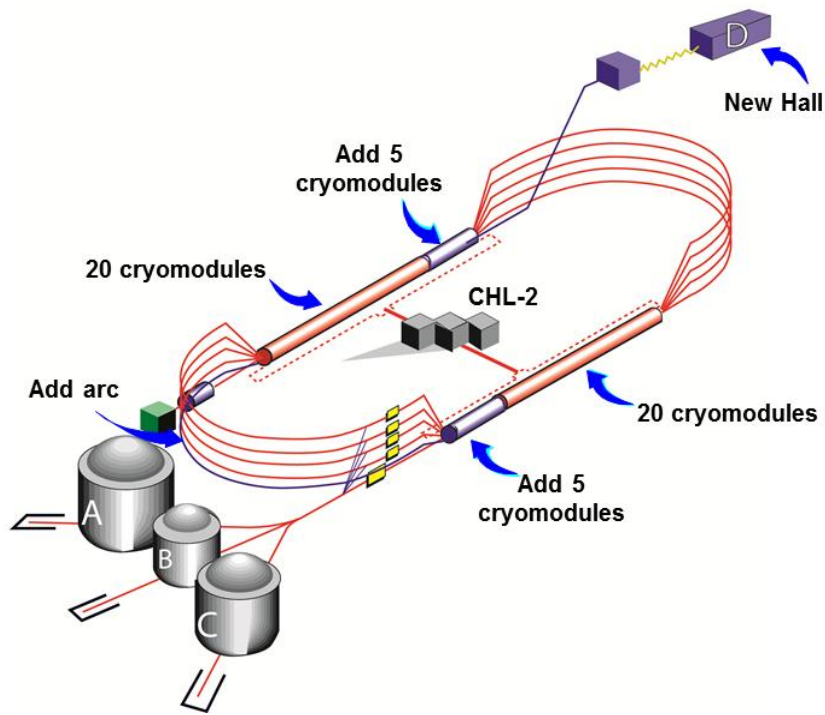

**Figure 1:** A sketch indicating the scope involved in the Jefferson Lab CEBAF 12 GeV Upgrade Project.

The accelerator upgrade was formally completed during the summer of 2014 and in the intervening period has established increasingly demanding performance as it has enabled commissioning of the GlueX experiment, execution of two experiments in Hall B "underneath" the CLAS12 construction, and operated at 5 pass, 10.6 GeV and below, for a combined magnetic form factor measurement and Deeply Virtual Compton Scattering experiment. The accelerator has been operated with ~750 kilowatts of beam power to Hall A. Halls A and D have been operated concurrently. A new 750 MHz separator system to enable concurrent operation to all four halls is in commissioning.

Hall D houses the GlueX experiment. The photon beam is generated using a diamond radiator which creates linearly polarised photons which enhance the differentiation between the multiple meson resonances, among which the experiment will seek the hybrid and exotic states predicted by Quantum ChromoDynamics, but never definitively observed. This physics is related to the searches for pentaquark and tetraquark systems which have recently re-energized the field with putative observations at the high energy colliders. The experiment is operational.

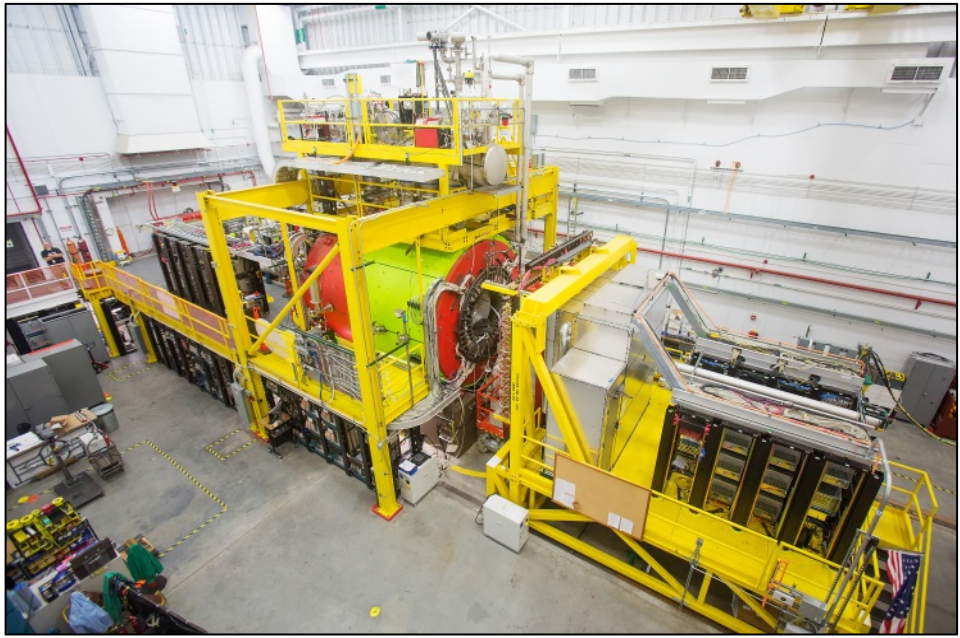

**Figure 2.** The GlueX Experiment in Hall D at Jefferson Lab.

Hall B will house the CEBAF Large Aperture Spectrometer (CLAS12). While similar in concept to the pre-existing CLAS spectrometer, the large TORUS magnet which recently reached full current, and the Solenoid magnet, which will encase the target, are both new as are essentially all of the associated drift chambers, silicon vertex tracker, Micromegas detectors, Cherenkov detectors and calorimeters. All the detectors are complete and being installed.

In Hall C, the High Momentum Spectrometer, matched to the top energies of the 6 GeV era is complemented by the Super High Momentum Spectrometer. In all, this spectrometer comprises five custom superconducting magnets, in order, a "septum" dipole to enable operation at very small scattering angles, three quadrupoles and a large bending dipole. All are mounted on the spectrometer carriage and all detectors are complete and mostly installed.

The experimental apparatus in both of Halls B, and C, will be commissioned and operational during 2017.

**2.    The 12 GeV Program**

We can express some high level goals of the 12 GeV program in terms of key questions:

1. What is the role of gluonic excitations in the spectroscopy of the light mesons?

2. Where is the missing spin in the nucleon?
    a. What is the role of orbital angular momentum of the partons?

3. Can we reveal a novel landscape of nucleon sub-structure through 3D imaging at the femtometer scale?

4. Can we discover evidence for physics beyond the standard model?

The Program Advisory Committee works with a somewhat more detailed set of categories:

A. The Hadron spectra as probes of QCD
   (GlueX and heavy baryon and meson spectroscopy)
B. The transverse structure of the hadrons
   (Elastic and transition form factors)
C. The longitudinal structure of hadrons
   (Unpolarized and polarized parton distribution functions)
D. The 3D structure of the hadrons.
   (Generalised Parton Distributions and Transverse Momentum Distributions)
E. Hadrons and cold nuclear matter.
   (Medium modifications of the nucleons, quark hadronization, N-N correlations, few-body experiments)
F. Low energy tests of the standard model and fundamental symmetries.

While the apparatus in the different halls has been designed to be complementary, it turns out that nearly all the topics have attracted proposals from several halls. For example, the program of experiments with the collective goal of a comprehensive study of the transverse momentum distributions comprises experiments from all of Halls A, B, and C. The program will fill a decade with top-of-the-line experiments.

**3.     Meson and Baryon Spectroscopy, Hybrid Mesons, and Confinement**

This is an entirely new thrust in physics for Jefferson Lab to make definitive studies of the light meson spectrum [2]. The GlueX experiment, with detailed partial wave analysis of the photoproduction data, will seek the numerous states which follow naturally in QCD but which have yet to be established. These studies bear sharply on the confinement of quarks and gluons within hadrons and ultimately within nuclei. The lattice Quantum Chromodynamics calculations, as illustrated in Figure 3, are carried out by a collaboration led by Jefferson Lab physicists and provide on the one hand, clear predictions, and, on the other hand, a strong framework for the high level analysis of the data. The Joint Physics Analysis Centre anticipates a strong collaboration with the experimental teams, GlueX and those at other laboratories to help elucidate the results.

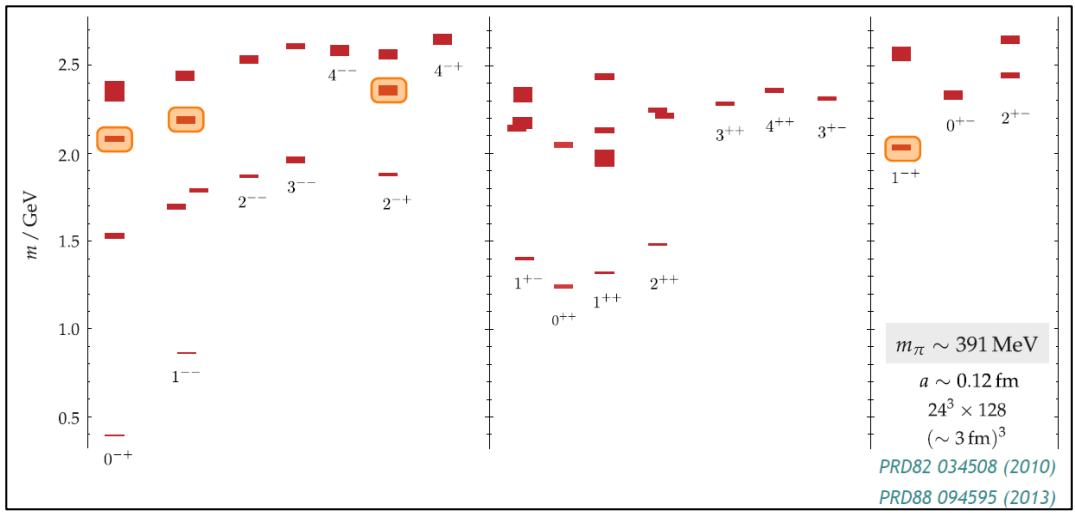

**Figure 3.** The light meson spectrum from recent lattice QCD calculations by the Hadron Spectrum Collaboration. The shaded states appear to form a hybrid super-multiplet $(0, 1, 2)^{-+}$, $1^{--}$ with mass about 0.9 GeV above that of the $\rho$ meson.

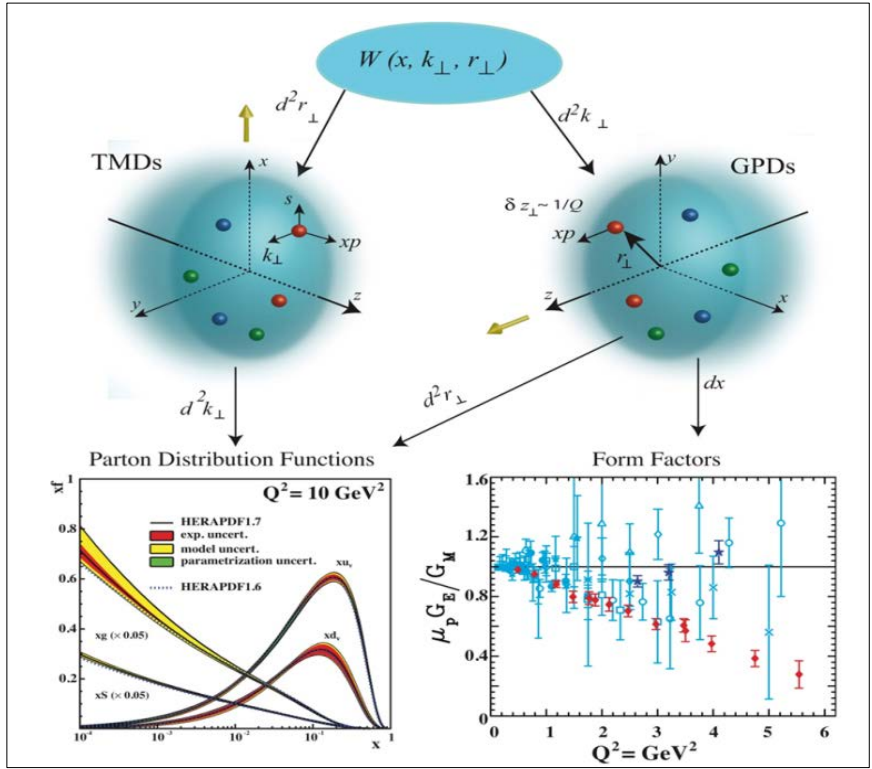

**Figure 4.** Illustration of the relationship between Transverse Momentum Distributions, Generalised Parton Distributions, Form Factors, and Parton Distribution Functions [1].

Beyond the GlueX experiment, which is anticipated to take several years to exhaust its potential, there are several ideas for further exploitation of the Hall D facility in the realm of, for example, rare η-meson decays [3], and $K^0_L$ physics [4].

**4.     Nucleon Tomography**

For five decades we have sought to understand how charge and magnetism are distributed in the nucleon or the nucleus. Since the 1970's we have related these form factors, and the longitudinal momentum, $x_{Bj}$, distributions to the disposition of partons. We are at the threshold of an era in which we probe this phase space in 3-dimensions of quarks and gluons within the target. Since the early '90s the appreciation of these distributions has deepened as we have related them to an underlying 5-dimensional quantum mechanical phase space of the partons. This is illustrated in Figure 4. Today, we explore the momentum space by measuring the transverse momentum distributions (TMD). In these measurements we integrate over the spatial distributions. Analogously, by integrating over the transverse momentum components, the Generalized Parton Distributions (GPD) measured in Deeply Virtual Compton Scattering provide a multi-dimensional spatial view of the nucleon. Thus, by dint of improved understanding of what we are doing, and by improved experimental techniques we hope to open a whole new vista of the micro-world.

**5.     Nuclei and Nuclear Structure**

While the mainstream of physics with CEBAF has been dominated by what is usually called "medium energy physics", some of the experiments have revealed important characteristics of the structure of nuclei. A clear example has been the relatively recent observation of strong p-n correlations among the nucleons emitted in deep inelastic scattering. These measurements are a rather clear demonstration of the characteristics of the short range correlations generated by the forces between nucleons in the nuclear potential.

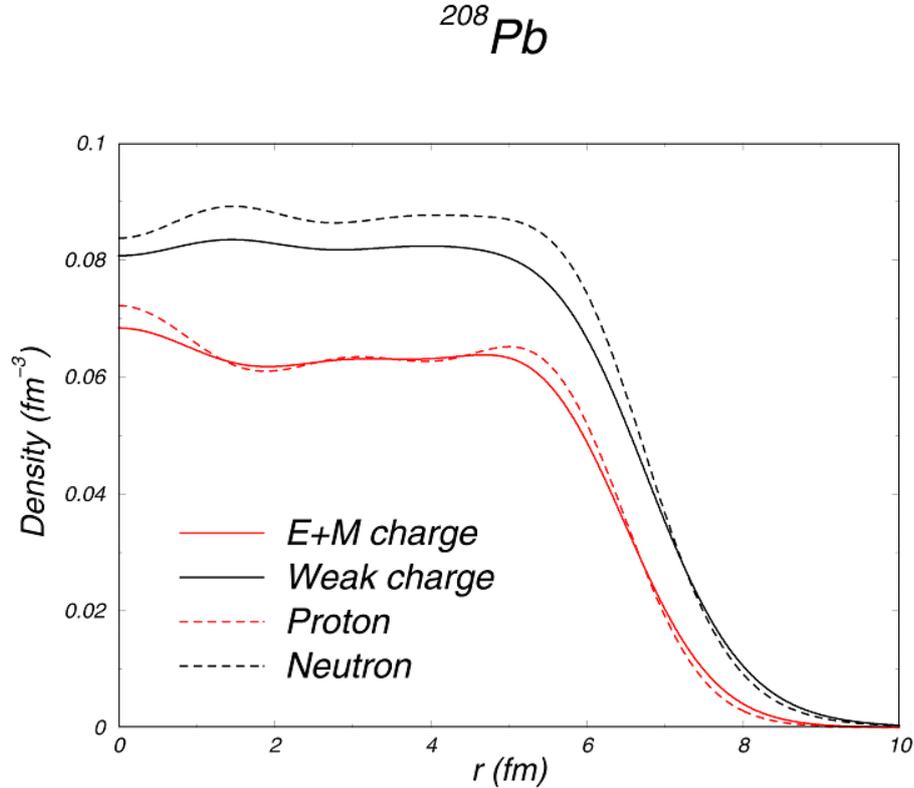

**Figure 5.** Figure showing the differences between the proton and neutron radial distributions in lead and the consequent distributions of electromagnetic and weak charges.

Parity violating electron scattering measurements are enabled by the exquisite control of the helicity of the electrons in CEBAF. The experiments can choose to flip the spin at frequencies of 960 Hz, thus providing excellent control of the systematic differences between the measurements made with the two states, and permitting sensitivities into the 10 parts per billion range. The weak coupling of the proton, $(1- \sin^2\theta_W)$, is very small since $\sin^2\theta_W \sim 0.25$. In contrast that of the neutron is unity. Therefore parity violation measurements are sensitive to the distribution of neutrons in the nucleus. If we take the case of Lead, one can relate the radius measurements to the equation of state of neutron stars. An initial measurement was published [5] by the PRex experiment. As part of the 12 GeV program, the lead radius measurement will be improved and among the obviously extensive set of nuclei for which such measurements would be interesting, measurements with Calcium targets, $^{40}$Ca and $^{48}$Ca, have been approved [6].

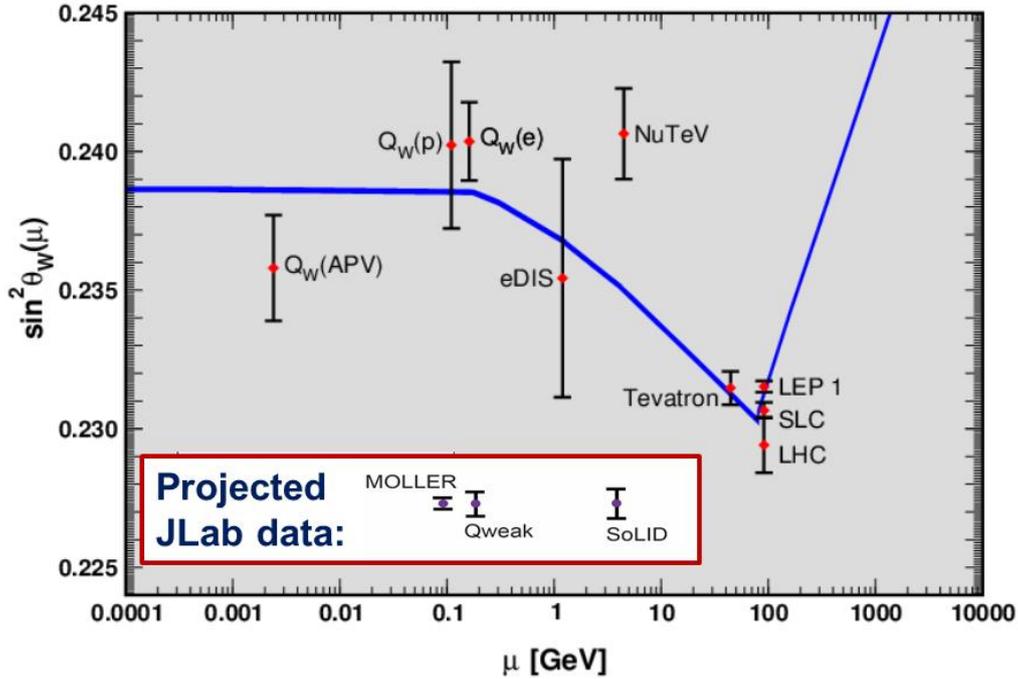

**Figure 6.** The weak coupling constant, $\sin^2\theta_W$, as a function of momentum transfer $\mu$. The anticipated uncertainties from measurements planned at Jefferson Lab are indicated [1].

## 6. The Standard Model and Beyond

The same parity violation techniques which were discussed above provide opportunities to make sensitive measurements of $\sin^2\theta_W$ itself. Such measurements have provided the foundations, along with the mass of the W boson, and the mass of the top quark, of the stress tests of the standard model. The most precise measurements of $\sin^2\theta_W$ were made by the SLC and LEP experiments. Thus far the low momentum transfer measurements have not been competitive, but that is changing. Figure 6 shows the current status of measurements. Jefferson Lab measurements of Qweak [7] of the proton and of the Deep Inelastic Scattering measurements (eDIS) [8] came from the later stages of the CEBAF 6 GeV running. The full data set for Qweak will significantly improve the precision, and the future measurements from SoLID [9], and especially MOLLER [10] will result in precision comparable to that from the high energy $e^+e^-$ colliders. Because the measurements at the high energies are dominated by the local Z-pole, the reach of the lower energy measurements is very good, extending into the TeV range. The Moller experiment is seeking CD0 (mission need), the first step project approval from the Department of Energy.

The bremsstrahlung process in the presence of a nucleus is a well-known method for producing photons by scattering electrons. This also would work for any other object with the same quantum numbers as the photon. During the last several years, prompted by some cosmic ray results, the possibility that a massive (but small mass) vector boson, a heavy photon, if you prefer might be the origin of dark matter. Three experiments have been proposed and received approval from Jefferson Lab. The APEX experiment searches for a peak in the mass distribution of electron-positron pairs using the two spectrometers in Hall A, has taken it first data and published limits [11]. The second, the Heavy Photon Search [12] operates a silicon detector close to the beam. This should enhance the sensitivity of the search by differentiating long lived decays. The experiment has had two beam periods, and we expect the results from the 2015 running soon. Finally, the Darklight [13] experiment plans to use the Low Energy Recirculator

Facility to search at very low energies with a very intense beam and a hermetic detector. Tests with a prototype experimental setup have taken place during the summer of 2016.

These experiments are further instances in which the properties of the beams available at Jefferson Lab enable experiments which broaden dramatically the potential 12 GeV program.

## 7. Conclusions

The 12 GeV Project is near to completion and elements of the physics program are established and operational in Halls A and D. The detectors for Halls B and C are in their final phases of construction and will be commissioned during 2017. The Physics program will be an exciting and perhaps dominant component of the Medium Energy Physics worldwide during the next decade. It is a very exciting time at Jefferson Lab.

## 8. Acknowledgements

The 12 GeV Program at Jefferson Lab has taken many years from its inception in the previous millennium to its realization as an important pillar of the world nuclear physics program in the 21st Century. I would like to acknowledge the input and contributions from the staff and users of Jefferson Lab who have made this exciting program possible. In particular, much of the material was shown by Bob McKeown at *Hadron Workshop 2016* in China, and some of the output from the GlueX experiment was presented by Curtis Meyer at *Meson 2016*.